\documentclass[twocolumn,showpacs,amsmath,amssymb,amsfonts,a4paper,aps,prd]{revtex4-1}
\usepackage{graphicx,color}\usepackage[linkbordercolor={1 0 0},citebordercolor={0 1 1}]{hyperref}\usepackage{float}\usepackage[dvipsnames]{xcolor}\usepackage{amsmath}
\usepackage{multirow}
\usepackage{float}
\usepackage[utf8]{inputenc}

\begin{document}

\title{ Power spectra in warm G-inflation and its consistency:  Stochastic approach}

\author{Meysam Motaharfar}
\email{mmotaharfar2000@gmail.com}
\affiliation{Department of Physics, Shahid Beheshti University, G. C., Evin,Tehran 19839, Iran}
\author{Erfan Massaeli}
\email{erfan.massaeli@gmail.com}
\affiliation{Department of Physics, Shahid Beheshti University, G. C., Evin,Tehran 19839, Iran}
\author{Hamid Reza Sepangi}
\email{hr-sepangi@sbu.ac.ir}
\affiliation{Department of Physics, Shahid Beheshti University, G. C., Evin,Tehran 19839, Iran}

\begin{abstract}
Recently, it has been realized that the so-called G-inflation model inspired by supplementing a generalized covariant Galileon-like non-linear derivative self-interacting term to the standard kinetic term should be ruled out from inflationary models. This is due to the fact that it suffers from lack of an oscillatory phase at the end of the inflationary regime which is typically accompanied by the appearance of a negative squared propagation speed  of the scalar mode leading to instabilities of small-scale perturbations. In this regard, the warm G-inflation scenario is proposed where for G-inflation to survive, the Galileon scalar field is coupled to the radiation field through a dissipation term which results in removing the reheating period due to the characteristics of warm inflationary scenario. In so doing, a linear stability analysis is first performed to obtain the appropriate slow-roll conditions in such a proposal. Cosmological perturbations of the model are then investigated by utilizing fluctuation-dissipation theorem and analytical expressions are derived for observable quantities; the power spectrum, tilt spectral index and tensor-to-scalar ratio in terms of $PSR$ parameters and Galileon flow functions. Finally, the model is solved for chaotic self-interacting potentials, particularly the renormalizable Higgs  potential $\frac{\lambda}{4} \phi^{4}$, and shown to be consistent with observations in the weak dissipation $Q \ll 1+ 3\frac{\delta_{GX}}{\delta_{X}}$ and G-dominant $3\frac{\delta_{GX}}{\delta_{X}}\gg 1$ regime despite its large self-coupling, since the energy scale at the horizon crossing is depressed by the synergy of Galileon and thermal effects.
\end{abstract}
\date{\today}
\maketitle
\section{introduction}

\textit{Inflation} \cite{Starobinsky:1979ty,Linde:1984ir}, a quasi-de-Sitter accelerating expansion phase which is realized by means of microphysical models including a dynamical field, the ``inflaton,'' evolving under the influence of a plateau-like potential ($H \propto \sqrt{V} \propto 10^{16}$ GeV), resolves a number of long-standing problems which the Standard Big Bang (SBB) cosmology is fraught with, such as the observed flatness, horizon, homogeneity and unwanted relics, to name but a few
\cite{Guth:1980zm, Linde:1981mu}.
The noteworthy feature of such an elegant paradigm is that it serves as a casual mechanism to seed the acoustic peaks in Cosmic Microwave Background (CMB) radiation as well as account for the distribution of Large-Scale Structure (LSS) from the evolution of primordial quantum vacuum fluctuations during inflation
\cite{Mukhanov:1981xt,Guth:1982ec,Starobinsky:1982ee, Bardeen:1983qw}.

The scenario in which inflaton is isolated and the interaction between inflaton and other subdominant fields is neglected whereby the universe undergoes a first order phase transition \cite{Starobinsky:1982ee} and its temperature drastically diminishes, is conventionally called cold inflation (isentropic). After such steep supercooling phase, the universe should go through a reheating phase \cite{Kofman:1994rk,Kofman:1997yn,Albrecht:1982mp}; oscillations of inflaton around the minimum of its potential in order to heat up again and progressively proceed to the radiation era required by SBB. While meshing these two isolated stages, the inflationary phase and the subsequent reheating phase, often brings about a severe discrepancy.
Therefore, warm inflation \cite{Berera:1995wh,Berera:1995ie} (non-isentropic), as a complementary scenario, has been constructed to avoid such problems by introducing a supplementary viscose term having a dissipation coefficient which illustrates the rate of energy exchange between inflaton and radiation field.
In fact, inflaton concurrently dissipates into radiation whereby primeval radiation will not heavily be diluted during inflation and smoothly enters the radiation era, for details see \cite{Yokoyama:1998ju,Taylor:2000ze,DeOliveira:2001he,DeOliveira:2002wk,Hall:2003zp,Moss:2007cv,Moss:2008yb,Berera:2008ar,Graham:2009bf,moss2011non,Bartrum:2013oka}. As a result, warm inflation not only inherits the features of conventional inflation but also removes disparities coming from the reheating phase and thus alleviates the initial condition
\cite{Berera:2000xz},
cures the overlarge amplitude of the inflaton field and circumvents the so-called eta-problem \cite{Berera:2004er}.
Furthermore, it contributes a very appealing mechanism for baryogenesis where spontaneous lepto/baryogenesis can easily be realized
\cite{Brandenberger:2003kc}.
In addition, the nature of fluctuations stems from thermal fluctuations in radiation which are coupled to inflaton due to the presence of dissipation coefficient rather than quantum fluctuations where the condition for which thermal fluctuations dominate over quantum fluctuations is given by $T>H$ \cite{Berera:1995wh, Berera:1995ie}. Moreover, the matter components of the universe are created by the decay of either the remaining inflaton field or the dominant radiation field.

From a quantum field theory perspective, the only known scalar field to drive inflation is the Standard Model (SM) Higgs boson. However, Higgs-driven
inflationary models with the standard kinetic term and renormalizable self-interaction potential \cite{Linde:1983chaotic} produce a large amplitude for the curvature and tensor perturbations which are not consistent with the observed universe \cite{Ade:2015xua}. To reconcile Higgs-driven inflationary models with observations, numerous modifications have been imposed to the effective Lagrangian in order to suppress the energy scale of inflation including a non-minimally coupled term to gravity with a large coupling \cite{Barvinsky:2008ia}, non-minimal coupling to the Higgs kinetic term with Einstein tensor \cite{Germani:2010gm,Germani:2010ux}, non-standard higher order kinetic term, dubbed k-inflation \cite{Armenda:1999rj,Garriga:1999vw} such as ghost condensate \cite{ArkaniHamed:2003uz} and Dirac-Born-Infeld inflationary models \cite{Alishahiha:2004eh}.
Apart from the first case which suppresses the energy scale of inflation by large effective Planck scale, the others are kinetically modified which means that extra viscosity terms have been added to dynamical equations  whereby the evolution of inflaton may be controlled and become consistent with observations even for large self-couplings and steeper potentials.

Incorporating higher order kinetic terms often lead to a new degrees of freedom followed by unwanted ghost instabilities. It would therefore be of interest to see if a scalar field, in spite of its higher derivative nature, does not result in a new degree of freedom.
Currently, it has been demonstrated that a particular combination of higher derivative kinetic terms not only maintains both the scalar and gravitational field equations to second order but also does not lead to new degrees of freedom \cite{Deffayet:2009wt,Deffayet:2009mn}. The scalar field having such properties is known as the Galileon since it possesses a Galileon shift symmetry in Minkowski background. Such a scalar field has initially been investigated in the context of modified gravity and dark energy \cite{Chow:2009fm,Silva:2009km}. Recently, a Galileon driven inflationary model dubbed G-inflation was proposed in \cite{Kobayashi:2010cm}.
The striking characteristics of such inflationary models are that they can produce scale-invariant spectral index even in an exactly de Sitter background and the tensor-to-scalar ratio can take larger values than that in conventional inflation due to the violation of the consistency relation, that is $r = - 8.7 n_{t}$. Although Higgs G-inflation \cite{Kamada:2010qe} is consistent with observations even for large self-coupling, which is roughly around $0.13$ from quantum field theory point of view,  it has very recently  been realized that it suffers from the absence of an oscillatory phase typically accompanied by a negatively squared propagating sound speed leading to a Laplacian equation for curvature perturbations instead of a wave equation, while producing unstable small-scale perturbations \cite{Ohashi:2012wf}. To resolve the problem, the authors in \cite{Kamada:2013bia} have added an extra quadratic non-standard kinetic term to the action in order to obtain positive sound speed resulting in the required reheating phase despite large self-coupling of the Higgs self-interaction potential.

Taken together, the incentive to investigate G-inflation in the context of warm inflation is to eradicate the aforementioned problems by  avoiding the reheating phase which is due to the nature of the warm inflationary scenario to survive G-inflation, particularly the Higgs G-inflation. Therefore, the layout of the paper is the following. We present the slow-roll dynamical field equations for warm G-inflation model, taking into account generalized Galileon scalar field through an arbitrary function of $\phi$ and $X$ as $G(\phi, X)$ and also temperature dependence of the potential and dissipation coefficient in \ref{2}. Next, the validity of slow-roll conditions is investigated by means of stability analysis applied to the dynamical system in section \ref{3}. In section \ref{4}, the cosmological perturbations are investigated utilizing fluctuation-dissipation theorem where the corresponding power spectrum, tilt spectral index and tensor-to-scalar ratio are also calculated in a generic form for the dissipation coefficient and potential. We then solve the model at hand for chaotic potentials, particularly Higgs boson self-interaction potential $\frac{\lambda}{4} \phi^{4}$ in section \ref{5}. Finally, the conclusion is drawn in section \ref{6}. Throughout the paper, we adopt the metric signature $ (-, +, +, +)$.

\section{Slow-roll regime of warm G-inflation}\label{2}

We begin with the multi-component, kinetically modified, minimally coupled action as follows
\begin{align}\label{action}
S = \int d^{4} x \sqrt{ - g} \Bigg[&
\frac{M^{2}_{pl}}{2} R + X- V(\phi, T) \notag\\& - G(\phi, X) \Box \phi + \mathcal{L}_{r} + \mathcal{L}_{int} \Bigg],
\end{align}
where $g$ is the determinant of the metric tensor $g_{\mu \nu}$, $R$ is Ricci scalar, $M_{Pl}= ({8 \pi G_{N}})^{-\frac{1}{2}} = 2.44 \times 10^{18}$ GeV is the reduced Planck mass with $G_{N}$ being the gravitational constant and $G(\phi, X)$ represents an arbitrary function of the scalar field $\phi$ and the standard kinetic term $X=-\frac12 g^{\mu \nu} \partial_{\mu} \phi \partial_\nu \phi$. Also, $\mathcal{L}_{r}$ and $\mathcal{L}_{int}$ denote the Lagrangian of the radiation field and  interaction term between inflaton and other subdominant fields, respectively.

Varying the action (\ref{action}) with respect to the metric, the total energy-momentum tensor $T_{\mu\nu}$ containing the inflaton and radiation field contributions read
\begin{align}
\nonumber T_{\mu\nu}& = \nabla_{\mu}\phi\nabla_{\nu}\phi+ g_{\mu\nu}\left(X - V(\phi, T)\right)-2\nabla_{(\mu}G\nabla_{\nu)}\phi\\&+ g_{\mu\nu}\nabla_{\lambda}G\nabla^{\lambda}\phi- G_{,X}\Box\phi\nabla_{\mu}\phi\nabla_{\nu}\phi + T s u_\mu u_\nu,
\end{align}
where $T$ is the universal temperature, $s$ is the entropy density and $u_{\mu} = (-1, 0, 0, 0)$ is the radiation fluid 4-velocity vector. Note that here and hereafter we utilize the notation $G_{,X}$ for $ \frac{\partial G}{\partial{X}}$. Taking variation of the action with respect to the scalar field also yields the following modified Klein-Gordon equation
\begin{align}\label{G1}
\nonumber&-\Box\phi+2\left(G_{,\phi}-G_{,\phi X}X\right)\Box \phi- G_{,X}\Big[\left(\nabla_{\mu}\nabla_{\nu}\phi\right)\left(\nabla^{\mu}\nabla^{\nu}\phi\right)\\&\nonumber-\left(\Box \phi\right)^{2}+ R_{\mu\nu}\nabla^{\mu}\phi\nabla^{\nu}\phi\Big]- 2 G_{,\phi X}\left(\nabla_{\mu}\nabla_{\nu}\phi\right)\left(\nabla^{\mu}\phi\nabla^{\nu}\phi\right)\\&\nonumber+
G_{,XX}\left(\nabla^{\mu}\nabla^{\lambda}\phi-g^{\mu\lambda}\Box\phi\right)\left(\nabla_{\mu}\nabla^{\nu}\phi\right)\left(\nabla_{\nu}\phi\nabla_{\lambda}\phi\right)+V_{,\phi}\\&- 2 G_{,\phi\phi}X=-\Gamma u^{\mu}\partial_{\mu}\phi.
\end{align}
Here, $R_{\mu\nu}$ denotes the Ricci tensor and $\Gamma$ is defined as a dissipation coefficient implying the rate of energy exchange between inflaton and radiation (i.e. the rate of inflaton decaying to radiation). Let us proceed by adopting a spatially flat Friedmann-Lemaitre-Robertson-Walker (FLRW) space-time with  scale factor $a(t)$ where $t$ is the cosmic time. Therefore, equation of motion  takes the following form
\begin{align} \label{a1}
\mathcal{B}\ddot\phi + 3 H \mathcal{A}\dot \phi+V_{,\phi}=0,
\end{align}
with
\begin{align}
\nonumber\mathcal{A}= &1+Q+ 3H\dot\phi G_{,X}+ \frac{\dot H \dot\phi G_{,X}}{H}+2XG_{,\phi X}- 2G_{,\phi} \\&- \frac{G_{,\phi\phi}\dot\phi}{3H}\\
\mathcal{B}= &1+6H\dot\phi G_{,X}+ 6H \dot\phi XG_{,XX}-2XG_{,\phi X}-2G_{,\phi},
\end{align}
where the dimensionless parameter $Q=\frac{\Gamma}{3 H}$ quantifies the strength of dissipation.
The energy-momentum tensor has the form $T^{\mu}_{ \nu}=\mbox{diag}\left(-\rho, p, p, p\right)$ with
\begin{align}
\rho& = X+V(\phi, T)+ 6HG_{,X}X\dot\phi-2G_{,\phi}X+ Ts,\\
p&= X-V(\phi, T) - 2\left(G_{,\phi}+G_{,X}\ddot\phi\right)X,
\end{align}
where $\rho$ and $p$ are total energy density and pressure of the system, respectively. Here, $\rho$ and $p$ have an explicit dependence on the Hubble rate, therefore, gravitational field equations are given by
\begin{align}
3M^{2}_{pl}H^{2}= \rho, \ \ \ \
-M^{2}_{pl}(3H^{2}+2\dot H) = p.
\end{align}

Now, considering the thermodynamic relation $U = F + Ts$ where $U$ is the total energy and $F$ is total free energy, the free energy density will be
\begin{align}
f = X+V(\phi, T)+ 6HG_{,X}X\dot\phi-2G_{,\phi}X.
\end{align}
Indeed, through the definition of entropy in thermodynamics, $s(\phi, T)$ has following relation with potential
\begin{align}
s = - \frac{\partial f}{\partial T} = - V_{,T}(\phi, T).
\end{align}
Using conservation equation $\dot \rho + 3 H (\rho + p) = 0$ together with Eq. (\ref{a1}), entropy production equation reads
\begin{align}\label{a2}
T\dot s+ 3HTs= \Gamma\dot \phi^{2}.
\end{align}
When thermal correction to field masses is negligible, i.e. $b \ll \frac{Q}{\mathcal{A}}$ in the slow-roll regime as we will see in the next section, the effective potential can be written as $V(\phi, T) \simeq V(\phi) + V(T)$ and therefore $V(T)$ contributes to the energy density of the radiation field resulting in $\rho_{\gamma} = \frac{3}{4} Ts$. Equivalently, conservation Eq. (\ref{a2}) is given by
\begin{align}\label{ss}
\dot \rho_{r} + 4H \rho_{r} = \Gamma \dot \phi^{2}.
\end{align}

Now lets us define a dimensionless HSR parameter as
\begin{align}\label{g1}
\nonumber \epsilon_{H} \equiv& -\frac{\dot H}{H^{2}},
\\\notag=& \frac{1}{2} \frac{2X+6HG_{,X}\dot \phi X-4G_{,\phi}X-2G_{,X}\ddot \phi X+  Ts}{M^{2}_{pl} H^{2}},
\\=& \delta_{X}+ 3 \delta_{GX}-2 \delta_{G\phi}- \delta_{\phi} \delta_{GX}+ \frac{3Ts}{2V},
\end{align}
where
\begin{align*}
\delta_{X} \equiv \frac{X}{M^{2}_{pl}H^{2}}, \delta_{GX} \equiv \frac{G_{,X}\dot\phi X}{M^{2}_{pl}H}, \delta_{G\phi} \equiv \frac{G_{,\phi}X}{M^{2}_{pl}H^{2}}, \delta_{\phi} \equiv \frac{\ddot \phi}{H \dot\phi}.
\end{align*}
The second HSR parameter $\eta_{H}$ characterizing the relative size of $\epsilon$ is systemically defined as follows
\begin{align}
\eta_{H}\equiv& \frac{d \ln \epsilon_{H}}{d \ln a} = \frac{\dot \epsilon_{H}}{H \epsilon_{H}}.
\end{align}
Accordingly, inflation then takes place when condition $\epsilon_{H}<1$ is satisfied, implying $\ddot a>0$ and will terminate when $\epsilon_{H} =1$.
Therefore, $\mathcal{A}$ and $\mathcal{B}$ can be reformulated in terms of the newly defined parameters as
\begin{align}
\mathcal{A} &= 1+Q+ \left(3-\epsilon_{H}\right)\frac{\delta_{GX}}{\delta_{X}} +2\left(\lambda_{X}-1\right) \frac{\delta_{G\phi}}{\delta_{X}}\notag,\\&- \frac{\sqrt{2}}{3}\lambda_{\phi} \frac{\delta_{G\phi}}{\sqrt{\delta_{X}}},\\
\mathcal{B}& = 1+ 6 (1+ \kappa_{X})\frac{\delta_{GX}}{\delta_{X}} - 2 (1+ \lambda_{X}) \frac{\delta_{G\phi}}{\delta_{X}},
\end{align}
where $\lambda_{X}$, $\lambda_{\phi}$ and $\kappa_{X}$ are defined as
\begin{align*}
\lambda_{X} \equiv \frac{X G_{,\phi X}}{G_{,\phi}}, ~ \lambda_{\phi} \equiv M_{pl} \frac{G_{,\phi\phi}}{G_{,\phi}}, ~ \kappa_{X} \equiv \frac{XG_{,XX}}{G_{,X}}.
\end{align*}

The difficulty in solving dynamical equations of inflation in exact form often leads to introducing a set of slow-roll approximations where the logarithmic variation of the Hubble parameter with respect to the e-folding number should be very small, $\dot H\ll H^{2}$, and the leading derivative terms $\ddot \phi \ll H \dot \phi$ \& $\dot s \ll H s$ are neglected,  implying that the energy is dominated by the potential, inflaton is slowly evolving and radiation is quasi-statically produced. Consequently, the first condition $(\dot H\ll H^{2})$ implies that each terms in Eq. (\ref{g1}) should be small (such assumption is to be taken with care since the sum of all terms may also be  small)
\begin{align}
\{\delta_{X}, \delta_{GX}, \delta_{G\phi}, \delta_{\phi}\}\ll 1, \ \ \ \ \ \ Ts \ll V,
\end{align}
and as a result, slow-roll equations take the following form
\begin{align}
3 M^{2}_{pl} H^{2} & \simeq V\label{r},\\
3 H \mathcal{A} \dot \phi+ V_{,\phi} & \simeq 0 \label{e1},\\
Ts &\simeq Q \dot \phi^{2} \label{e2},
\end{align}
and $\mathcal{A}$ reduces to
\begin{align}
\mathcal{A} \simeq 1+Q+ 3\frac{\delta_{GX}}{\delta_{X}} +2\left(\lambda_{X}-1\right) \frac{\delta_{G\phi}}{\delta_{X}}.
\end{align}
Furthermore, the number of e-folding is defined as follows
\begin{align}\label{vv}
N \equiv \int^{t_{end}}_{t_{hc}} H dt = \int^{\phi_{end}}_{\phi_{hc}} \frac{H}{\dot\phi} d\phi = - \int^{\phi_{end}}_{\phi_{hc}} M^{-2}_{pl}\frac{V}{V_{,\phi}} \mathcal{A} d \phi,
\end{align}
where $\phi_{hc}$ denotes the value of the inflaton field at the Hubble crossing time and $\phi_{end}$ represents the value of the inflaton field at the end of inflation.

Before closing this section, in order to have a concise investigation of the consistency of warm G-inflation in the upcoming section we introduce the customary dimensionless PSR parameters as \cite{Liddle:1994dx}
\begin{align*}
\epsilon \equiv \frac{M^{2}_{pl}}{2} \left(\frac{V_{,\phi}}{V}\right)^{2}, ~ \eta \equiv M^{2}_{pl}\frac{V_{,\phi\phi}}{V}, ~ \beta \equiv M^{2}_{pl}\frac{V_{,\phi}\Gamma_{,\phi}}{V\Gamma},
\end{align*}
which are supplemented with two more parameters, namely
\begin{align}
b \equiv \frac{T V_{,\phi T}}{V_{,\phi}}, \ \ c \equiv \frac{T \Gamma_{,T}}{\Gamma},
\end{align}
which account for temperature dependence of the potential and damping coefficient, respectively.
\section{stability analysis}\label{3}

The slow-roll approximations which were used above and led to slow-roll Eqs. (\ref{r}--\ref{e2}), immediately bring up the question as to under what conditions can slow-roll equations portray the system well? To answer the question, let us consider the inflaton field as an independent variable for which dynamical Eqs. (\ref{a1}, \ref{a2}) can be expressed in terms of its first derivative, that is
\begin{align}\label{a3}
 u^{\prime}& = -\mathcal{B}^{-1}\left[3H\mathcal{A}+V_{,\phi} u^{-1}\right] = f(u, s, \phi),
\\\label{a4}
 s^{\prime}& = -3Hsu^{-1}+ {\Gamma u}{T^{-1}} = g(u, s, \phi),
\end{align}
where a prime denotes derivative with respect to $\phi$ and $u = \dot\phi$. A glance at Eqs. (\ref{a3}, \ref{a4} reveals that these equations form a two dimensional dynamical system where its fixed points can be obtained by equating $f$ and $g$ to zero
\begin{align}\label{d1}
u_{0} &= -\frac{V_{,\phi}}{3H\mathcal{A}},
\\\label{d2}
s_{0} &= Q u_{0}^{2} T^{-1},
\end{align}
where a subscript zero means that $u_{0}$ and $s_{0}$ are the solution of the so-called slow-roll Eqs. (\ref{e1}, \ref{e2}). To answer the question of validity of the slow-roll approximations involves a linear stability analysis which means perturbing exact solutions around the slow-roll solutions in order to see  under what conditions the system will remain close to the slow-roll solutions for many Hubble times. In other words, the slow-roll solutions should be the attractors of the dynamical system.
Therefore, perturbing variables $u$ and $s$ around the slow-roll solutions, that is $u \rightarrow u_0 + \delta u$ and $s \rightarrow s_0 + \delta s$, with perturbed terms being much smaller than the background ones ($\delta u_{0}\ll u_{0}$ \& $\delta s_{0} \ll s_{0}$), we have
\begin{align}
\delta x^{\prime} = M(x_{0})\delta x - x_{0}^{\prime},
\end{align}
with
\begin{equation}
x = \left(\begin{array}{c} u \\ s \end{array}\right),
\end{equation}
and $M$ is $2 \times 2$ matrix given by
\begin{align}
M = \left( \begin{array}{cc} A & B \\ C & D \end{array}\right) = \left( \begin{array}{cc} \frac{\partial f}{\partial u} ~&~ \frac{\partial f}{\partial s} \\ \\  \frac{\partial g}{\partial u}~&~ \frac{\partial g}{\partial s} \end{array}\right)_{\left|\begin{array}{c} u=u_0 \\ s =s_0\end{array}\right.}.
\end{align}
The matrix elements of $M$ are
\begin{align}
A&= \frac{H}{u\mathcal{B}}\bigg( - 3 \mathcal{A} - 3 \left( 2 \kappa_{X}+1 \right) ( 3 - \frac{\epsilon}{\mathcal{A}}) \frac{\delta_{G X}}{\delta_{X}}
\notag\\&- \left( \frac{\epsilon}{\mathcal{A}^{2}} + 3 \left( 2 \kappa_{X}+3 \right)\delta_{G X}-2\left(\lambda_{X}+1 \right) \delta_{G \phi}\right)
\notag\\& \times \left(1+6\frac{\delta_{GX}}{\delta_{X}}+2(\lambda_{X}-1)\frac{\delta_{G\phi}}{\delta_{X}}\right)
\notag -12 \lambda_{XX} \frac{\delta_{G \phi}}{\delta_{X}}
\notag\\& + 2 (\kappa_{\phi\phi} + 3)\delta_{G X}+ 9 (2\kappa_{X}+\frac{\mathcal{A}}{\mathcal{B}}+3) \frac{\delta^{2}_{GX}}{\delta_{X}}
\notag\\&+ \sqrt{2} \lambda_{\phi} \frac{\delta_{G\phi}}{\sqrt{\delta_{X}}}-12 (\lambda_{X}+1) \frac{\delta_{G\phi} \delta_{G X}}{\delta_{X}}\notag\\& +9\frac{\delta_{G X}^2}{\mathcal{B}\delta_{X}^2}\left(3(2 \kappa_{X}+1)\delta_{GX} + 4 \lambda_{XX} \delta_{G\phi}\right)
\bigg),\label{matrixelementA}\\
\nonumber B&= \frac{H}{s \mathcal{B}}\Big( - \frac{Q\epsilon}{\mathcal{A}^{2}}\left(1+6\frac{\delta_{GX}}{\delta_{X}}+2 (\lambda_{X}-1) \frac{\delta_{G\phi}}{\delta_{X}}\right)  \\&+ 4 Q \delta_{GX}  -\left(1- \frac{3}{\mathcal{B}} \frac{\delta_{GX}^{2}}{\delta_{X}}\right)c Q + \mathcal{A} b \left(1- \frac{3}{\mathcal{B}} \frac{\delta_{GX}^{2}}{\delta_{X}}\right)\Big)\label{matrixelementB},
\\C&= \frac{Hs}{u^{2}} \Big(6 - \frac{\epsilon}{\mathcal{A}^{2}}- 3\left(2\kappa_{X}+3\right)\delta_{GX}
+2\left(\lambda_{X}+1\right)\delta_{G\phi}\Big)\label{matrixelementC},
\\D&= \frac{H}{u}\left(c - 4 - \frac{Q \epsilon}{\mathcal{A}^{2}}\right)\label{matrixelementD}.
\end{align}
These matrix elements reduce to the corresponding results obtained in \cite{Moss:2008yb} for $G = 0$. To derive Eqs. (\ref{matrixelementA}--\ref{matrixelementD}), we have utilized the following useful expressions
\begin{align}
3u\frac{H_{,u}}{H} \simeq &\frac{\epsilon}{\mathcal{A}^{2}}+ 3\left(2\kappa_{X}+3\right)\delta_{GX}-2\left(\lambda_{X}+1\right)\delta_{G\phi},
\end{align}
\begin{align}
-6 X G_{,X}\frac{\dot H_{,u} }{H} &\simeq9~ \frac{\delta_{G X}^2}{\mathcal{B}\delta_{X}^2}\Big(3(2 \kappa_{X}+1)\delta_{GX} + 4\lambda_{XX} \delta_{G\phi}\Big)
\notag\\&+\nonumber9 (2\kappa_{X}+\frac{\mathcal{A}}{\mathcal{B}}+3) \frac{\delta^{2}_{GX}}{\delta_{X}} +6\delta_{GX} \\& -12 (\lambda_{X}+1) \frac{\delta_{G\phi} \delta_{G X}}{\delta_{X}},
\end{align}
\vspace{-.2cm}
\begin{align}
3 \frac{s H_{,s}}{H} = \frac{Q\epsilon}{(1-\delta_{GX})\mathcal{A}^{2}} \simeq \frac{Q\epsilon}{\mathcal{A}^{2}}, ~ V_{,\phi s} = \frac{V_{\phi}b}{3s}, ~ \Gamma_{,s} = \frac{HcQ}{s}
\end{align}
\vspace{-.5cm}
\begin{align}
- 3 \frac{s \dot H_{,s} \dot \phi G_{,X}}{H} \simeq 4 Q \delta_{GX} + 3 \frac{c Q}{\mathcal{B}}\frac{\delta^{2}_{GX}}{\delta_{X}}- \frac{3\mathcal{A}b}{\mathcal{B}} \frac{\delta_{GX}^{2}}{\delta_{X}},
\end{align}
with $\lambda_{XX}$, $\lambda_{\phi\phi}$, and $\kappa_{\phi\phi}$ defined as follows
\begin{align*}
\lambda_{XX} \equiv \frac{X^{2}G_{,\phi XX}}{G_{,\phi}}, \lambda_{\phi\phi} \equiv M^{2}_{pl} \frac{G_{,\phi\phi\phi}}{G_{,\phi}}, \kappa_{\phi\phi} \equiv M^{2}_{pl} \frac{G_{,\phi\phi X}}{G_{,X}}.
\end{align*}

Being the attractor for a warm inflationary dynamical system only depends on the nature of the eigenvalues. To put it differently, the slow-roll solutions can be an attractor when the eigenvalues of the matrix $M$ are negative or possibly positive, but of order ${\epsilon}$ (i.e. it should be slowly evolving) \cite{Moss:2008yb} where $\epsilon$ refers to slow-roll parameters in general. More clearly, the determinant should be positive ($\mbox{det}  M>0$) and the trace should be negative ($\mbox{Tr} M<0$). Equally importantly, the force term $x_0^\prime$ whose size depends on the logarithmic derivative of $x_0$ with respect to e-folding number should be small enough (i.e. $\frac{\dot u}{H u}\ll 1$ and $\frac{\dot s}{H s}\ll 1$). Taking the time derivative of slow-roll Eqs. (\ref{d1}, \ref{d2}), we arrive at the leading order of $\epsilon$
\begin{align}\label{pp}
&\delta_{\phi} = \frac{\dot u_{0}}{H u_{0}} =\frac{1}{\Delta}\bigg\{\frac{4 Q}{\mathcal{A}}\beta+ (c-4)\eta + 3 \mathcal{A} b c
\notag\\& +\frac{\epsilon}{\mathcal{A}} \Big(4 - c(1+Q)+ \Big(\frac{1-2 c Q}{2 \mathcal{A}}\Big) \bigg[12 \frac{\delta_{G X}}{\delta_X}  + 4(2 \lambda_{X}-1) \frac{\delta_{G \phi}}{\delta_X}
\notag\\& - 12 ( 2 \kappa_{X} + 1 ) \frac{\lambda_{X}}{\mathcal{A} \sigma} \frac{\delta_{G \phi} \delta_{G X}}{\delta_X^2} \bigg]\Big)+ \Big(\frac12 - c Q\Big)
\notag\\& \times \bigg( - 4\kappa_{\phi\phi} \delta_{G X} + 4 \sqrt{2} \lambda_{\phi} \frac{\delta_{G \phi}}{\sqrt{\delta_X}} + \frac43 \lambda_{\phi\phi} \delta_{G \phi}
\notag\\ &- 8\frac{\lambda_{X} \kappa_{\phi\phi}}{\mathcal{A} \sigma} \frac{\delta_{G \phi} \delta_{G X}}{\delta_X^2} - 12 \sqrt{2} \frac{\lambda_{X} \lambda_{\phi}}{\mathcal{A}\sigma}\frac{\delta_{G \phi}^2}{\delta_X^{\frac32}} - 12\lambda_{X} \frac{\delta_{G \phi}}{\delta_X}\bigg)\bigg\},
\end{align}
\begin{align}\label{pp1}
&\delta_{s} = \frac{\dot s_{0}}{H s_{0}} = \frac{3}{\Delta}\bigg\{\frac{2 Q - \mathcal{A} \sigma}{\mathcal{A}} \beta + \frac{\mathcal{A} b}{Q} \Big(2 c Q + \mathcal{A} \sigma (1-c) \Big)
\notag\\& - 2 \eta+ \frac{\epsilon}{\mathcal{A}} \bigg( 2 + \mathcal{A} \sigma + 12 \frac{\delta_{G X}}{\delta_X} + 4(2 \lambda_{X}-1) \frac{\delta_{G \phi}}{\delta_X}
\notag\\& - 12 ( 2 \kappa_{X} + 1 ) \frac{\lambda_{X}}{\mathcal{A} \sigma} \frac{\delta_{G \phi} \delta_{G X}}{\delta_X^2} \bigg)-4 \kappa_{\phi\phi} \delta_{G X} + 4 \sqrt{2} \lambda_{\phi} \frac{\delta_{G \phi}}{\sqrt{\delta_{X}}}
\notag\\& + \frac43 \lambda_{\phi\phi} \delta_{G \phi} - 8 \frac{\lambda_{X} \kappa_{\phi\phi}}{\mathcal{A} \sigma} \frac{\delta_{G \phi} \delta_{G X}}{\delta_{X}^{2}}-12\sqrt{2} \frac{\lambda_{X} \lambda_{\phi}}{\mathcal{A} \sigma} \frac{\delta_{G \phi}^2}{\delta_X^{\frac32}}
 \notag\\&-12 \lambda_{X} \frac{\delta_{G \phi}}{\delta_X}\bigg\},
\end{align}
with $\Delta$ and $\sigma$ being defined as
\begin{align}
\Delta \equiv& (4-c) \mathcal{A}\sigma + 2cQ,
\\\sigma \equiv& 1+ \frac{1}{\mathcal{A}}\left(3\left(1+ 2 \kappa_{X}\right)\frac{\delta_{GX}}{\delta_{X}}+ 4 \lambda_{XX} \frac{\delta_{G\phi}}{\delta_{X}}\right).
\end{align}
We again note that Eqs. (\ref{pp}, \ref{pp1}) reduce to Eqs. (30) and (31) in \cite{Moss:2008yb} for $G=0$. The Hubble parameter should also be slowly varying, i.e. $\frac{\dot H}{H^2}=-\frac{\epsilon}{\mathcal{A}}\ll1$ which leads to a sufficient condition to satisfy the above requirement
\begin{align}\label{f1}
\{ |\epsilon|, |\eta|, |\beta| \} \ll \mathcal{A}, ~  0\le b \ll \frac{Q}{\mathcal{A}},  ~ \left|G_{,\phi}\right| = \left|\frac{\delta_{G\phi}}{\delta_{X}}\right| \ll 1.
\end{align}
However, the condition $\frac{\delta_{G\phi}}{\delta_{X}}\ll 1$ in weak dissipation $Q \ll 1+ 3\frac{\delta_{GX}}{\delta_{X}}$ and G-dominant  $3\frac{\delta_{GX}}{\delta_{X}}\gg 1$ regime translates into
\begin{align}
\left|\frac{\delta_{G\phi}}{3\delta_{GX}}\right| \ll 1.
\end{align}
Inspection of the above conditions shows that some terms in $\delta_{\phi}$ and $\delta_{s}$ will be of order $\epsilon^{2}$ and higher, and therefore we neglect them during calculations since they are too small. Also, $\sigma$  and $\mathcal{A }$ reduce to
\begin{align}
\sigma \equiv& 1+ \frac{1}{\mathcal{A}}\left(3\left(1+ 2 \kappa_{X}\right)\frac{\delta_{GX}}{\delta_{X}}\right),
\end{align}
and
\begin{align}
\mathcal{A} = 1+ Q + 3 \frac{\delta_{GX}}{\delta_{X}}.
\end{align}

In fact, conditions (\ref{f1}) generalize the standard slow-roll in the supercooled case including two over-damping terms, namely the thermal and Galileon friction terms which illustrate that inflationary potentials have broader choices whereupon  further novel inflationary models with steeper potential may be constructed in warm G-inflation. Furthermore, the last condition implies that the kinetic part of the Galileon scalar field interaction plays a substantial role in dynamics of the model at hand. In addition, the condition on slow-roll parameter $b$ implies that thermal corrections to the potential should be as small as in the absence of Galileon scalar field. As a result, the total energy density can be written in a separable form as $\rho(\phi, T) = \rho_{inf}(\phi)+ \rho_{rad}(T)$ which guarantees the former assumption.

So far, we have obtained conditions on the whole parameters of the model except for $c$. In order to obtain conditions  on $c$, we derive the determinant and trace of matrix $M$ at zero order of $\epsilon$ as follows
\begin{align}
\nonumber \det( M )&= \frac{H^{2}}{u^{2}\mathcal{B}}\left((c-4)\left(-3\mathcal{A}-9 (2 \kappa_{X}+1)\frac{\delta_{GX}}{\delta_{X}} \right) \right. \\& \left.+ 6 c Q - 6 \mathcal{A}b\right),\\
\mathrm{tr}(M)&=\frac{H}{u \mathcal{B}}\left(-3 \mathcal{A} - 9 (2 \kappa_{X}+1)\frac{\delta_{GX}}{\delta_{X}} + (c-4) \mathcal{B}\right),
\end{align}
where $\mathcal{B}$ reduces to
\begin{align}
\mathcal{B} = 1 + 6 (\kappa_{X}+1) \frac{\delta_{GX}}{\delta_{X}}.
\end{align}
Here, $\mathcal{B}$ should be positive since it later appears under a radical in power spectrum (\ref{qa}) and $\delta_{GX}$ should also be positive in order to avoid instability and ghosts \cite{De Felice,Avoidance}, therefore, $\kappa_{X}$  and $\mathcal{A}$ are positive quantities. Consequently, to have $\mbox{det}( M )>0$ and $\mathrm{Tr}(M)<0$, one arrives at the condition on $c$ as $$|c| \leq 4-2b,$$
and since $0\le b \le \frac{Q}{\mathcal{A}}$, we find
\begin{align}
|c|<4,
\end{align}
which means that temperature dependence of the dissipative coefficient should be within the range $\Gamma \propto (T^{-4}, T^{4})$. Utilizing Eq. (\ref{e2}) one deduces that $\frac{\rho_{rad}}{V} \simeq \frac{Q \epsilon}{2\mathcal{A}^{2}}$ which in turn means that the radiation field is sub-leading during the slow-roll regime which is consistent with the requirement for a warm inflationary scenario. Before closing the section, it is useful to derive some useful expressions which can also be used in the forthcoming section
\begin{align}
& \delta_{T} \equiv \frac{d \ln T}{d N}=\frac{\dot T}{H T}  = \frac{1}{3}\delta_{s} - \frac{\mathcal{A}b}{Q},
\\&\delta_{\Gamma} \equiv \frac{d \ln \Gamma}{d N} = \frac{\dot \Gamma}{H\Gamma} = -\frac{\beta}{\mathcal{A}} - \frac{\mathcal{A} b c}{Q} + \frac{c}{3} \delta_{s},
\\&\delta_{V}  \equiv \frac{d \ln V}{d N} = \frac{\dot V}{H V}  = -2 \frac{\epsilon}{\mathcal{A}} + \frac{2}{3} \frac{\epsilon}{\mathcal{A}} b - \frac{2Q}{9} \frac{\epsilon}{\mathcal{A}^{2}} \delta_{s},
\\&\delta_{V_{,\phi}}\equiv \frac{d \ln V_{,\phi}}{d N} = \frac{\dot V_{,\phi}}{H V_{,\phi}} = -\frac{\eta}{\mathcal{A}} - \mathcal{A} \frac{b^{2}}{Q} + \frac{1}{3}\delta_{s} b.
\end{align}
A glance at the above expressions reveals that  temperature, potential and the dissipative coefficient are all slowly varying parameters during slow-roll inflation which is consistent with the nature of inflation.

\section{Density fluctuations}\label{4}
It is now the time to develop the theory of cosmological perturbations in warm G-inflation. Since we are working in the context of a cosmological system,  metric perturbations  as well as field and thermal fluctuation should be included. As is well known, the prime characteristic of dissipating inflationary models which distinguishes them from the so-called cold inflation is that the nature of density fluctuations is due to thermal fluctuation in radiation field rather than in quantum fluctuations. These thermal fluctuations in radiation field are coupled to the inflaton field through the presence of damping terms in dynamical equations of inflation and their amplitude is fixed by the fluctuation-dissipation theorem. This means that both entropy and curvature perturbations must contribute to density fluctuations \cite{Moss:2008yb,Hall:2003zp}.

During inflation the energy density of radiation is subdominant, therefore, its thermal fluctuation merely contributes to entropy perturbations. Furthermore, in such system with a heat bath, entropy perturbations decay on scales larger than horizon and Consequently, one should keep track of curvature perturbations. Primordial cosmological perturbations are typically expressed in terms of  curvature perturbation on uniform energy density hyper surfaces denoted by $\mathcal{R}$. The reason behind using this quantity is that it is conserved on large scales in simple models, even beyond linear order  perturbation theory. In the linear order perturbation theory for the slow-roll single field inflation (warm-G-inflation is dominated by one single canonical field kinetically modified by the Galilean field interaction in over damped slow-roll regime) the curvature perturbation on the uniform density hypersurface is given by the gauge invariant linear combination $\mathcal{R}= \psi + \frac{H}{\dot \rho}\delta \rho $ with $\psi$ being the spatial metric perturbation and $\delta \phi$ representing perturbations about homogeneous inflaton field, respectively. For convenience, we choose the spatially flat gauge and accordingly, the perturbation is given by $\mathcal{R}=\frac{H}{\dot\rho}\delta\rho$ which in the slow-roll regime becomes $\mathcal{R} = \frac{H}{\dot\phi}\delta\phi$ \cite{ncanonical}. With these in mind, we expand the full inflaton field as $\Phi(t, x)=\phi(t)+\delta\phi(t, x),$ with $\phi(t)$ being the homogeneous background field and $\delta\phi$ a small perturbation $\delta\phi\ll\phi(t)$.

To compute the value of Fourier transformation of the inflaton fluctuation $\delta\phi$, a stochastic field approach is commonly utilized. In fact, the interaction between the inflaton field and radiation can be analyzed within the Schwinger-Keldysh approach in non-equilibrium field theory. Therefore, evaluation of fluctuations in expanding universe is obtained by applying equivalence principle to the non-expanding universe leading to the generalized second order Langevin equation after introducing stochastic thermal noise $\xi(t, x)$ \cite{Taylor:2000ze}
\begin{align}\label{G2}
 \mathcal{B}^{\prime} \ddot \Phi(x, t) &+ 3 H \mathcal{A} \dot \Phi(x, t) \notag\\ &+ V_{,\Phi} - \mathcal{F}\frac{\nabla^{2}}{a^{2}} \Phi(x,t) + \mathcal{K} (\Phi)= \xi(x,t),
\end{align}
where
\begin{align}
\mathcal{B}^{\prime} &= \mathcal{B} - 2 G_{,X} \frac{\nabla^{2}}{a^{2}}\Phi- 2 X G_{,XX} \frac{\nabla^{2}}{a^{2}}\Phi,\\
\mathcal{F} &= 1- 2 G_{,\Phi} + 2 G_{,\Phi X} + 4 H \dot \Phi G_{,X},
\end{align}
where $\mathcal{K}$ is a function of multiple terms of spatial derivative of $\Phi$ or its higher order derivative. Therefore, expanding the inflaton field the term $\mathcal{K}$ will at least be of order $(\delta \phi)^{2}$ which can be ignored and consequently, second order Langevin equation for the perturbed inflaton field in Fourier space is given by
\begin{align}
\nonumber \mathcal{B} \delta \ddot\phi(k, t) &+ 3 H \mathcal{A}^{\prime} \delta \dot \phi(k, t) + (\mathcal{F}^{\prime}\frac{k^{2}}{a^{2}}+V_{,\phi\phi}+ \mathcal{S}^{\prime}) \delta\phi(k,t) \\&= \xi(k,t),
\end{align}
with
\begin{align}
\nonumber \mathcal{A}^{\prime} &= \left(1+ Q - 2 \left(1-\lambda_{X} - 2 \lambda_{XX} \right) \frac{\delta_{G\phi}}{\delta_{X}}- \frac{2}{3} \lambda_{\phi\phi} \delta_{GX}\right. \\& \left.+ 2\left(3-\epsilon_{H}\right)\left(1+\kappa_{X}\right)\frac{\delta_{GX}}{\delta_{X}} - \frac{2\sqrt{2}}{3} \lambda_{\phi} \frac{\delta_{G\phi}}{\sqrt{\delta_{X}}}\right), \\
\nonumber \mathcal{S}^{\prime}  &= 3H^{2}\left(- \epsilon_{H}+Q \delta_{\Gamma}  +2 \lambda_{\phi\phi} \delta_{GX}-2 \sqrt{2} \lambda_{\phi} \frac{\delta_{G\phi}}{\sqrt{\delta_{X}}}\right.\\&\left.-\frac{2}{3} \kappa_{\phi\phi} \delta_{X} +2\left(3-\epsilon_{H}\right) \kappa_{X} \frac{\delta_{G\phi}}{\delta_{X}}\right),
\end{align}
\begin{align}
\mathcal{F}^{\prime} = 1 - 2 \left(1- \lambda_{X}\right)\frac{\delta_{G\phi}}{\delta_{X}} + 4 \frac{\delta_{GX}}{\delta_{X}} + 2 \frac{\delta_{\phi}\delta_{GX}}{\delta_{X}} \left(1+ \kappa_{X}\right).
\end{align}

To compute the power spectrum at the Hubble crossing point we note that Hubble crossing occurs well inside the slow-roll regime and as was discussed in stability analyses, the slow-roll regime is well consistent and therefore, the \textit{inertia} terms can be ignored. As a result, the first derivative Langevin equation in Fourier space takes the form
\begin{align}\label{q1}
3H\mathcal{C} \delta \dot\phi(k, t) + \left(\mathcal{D} \frac{k^{2}}{a^{2}}+ V_{, \phi \phi}\right) \delta \phi(k, t) = \xi (k, t),
\end{align}
with
\begin{align}
 \mathcal{C} &= \left(1+ Q + 6\left(1+\kappa_{X}\right)\frac{\delta_{GX}}{\delta_{X}}\right) = Q + \mathcal{B}, \\
 \mathcal{D} &=  \left(1 + 4 \frac{\delta_{GX}}{\delta_{X}}\right), \\&\nonumber
\end{align}
where we have neglected the terms first order in $\epsilon$ in the coefficient of perturbed inflaton since they will later  appear as second order in the spectral index. If the temperature in the universe is sufficiently high, the thermal noise is assumed to be Markovian and having the following properties
\begin{align}
\langle\xi(k,t)\rangle = 0,
\end{align}
\begin{align}
\langle\xi(k,t)\xi(-k^{\prime},t^{\prime})\rangle_{\xi} \overset{\underset{\mathrm{T \rightarrow \infty}}{}}{=} 2\Gamma T (2\pi)^{3} \delta^{3}(k-k^{\prime}) \delta(t-t^{\prime}).
\end{align}
The approximate analytical solution is
\begin{align}
\nonumber \delta \phi (k, t) &\approx \frac{1}{\mathcal{C}} e^{-(t-t_{0})/\tau(\phi)}\int^{t}_{t_{0}} e^{(t^{\prime}-t_{0})/\tau(\phi_{0})} \xi(k, t^{\prime}) dt^{\prime} \\ & +\delta\phi(k, t_{0}) e^{-(t-t_{0})/\tau(\phi_{0})}
\end{align}
where $\tau(\phi) = \frac{3H\mathcal{C}}{\mathcal{D} \frac{k^{2}}{a^{2}} +m^{2}}$  with $m^{2} = V_{,\phi\phi}$, represent the efficiency of thermalizing processes. The first term in the right hand side acts to thermalize $\delta \phi$, whereas the last is the memory term for the initial value of $\delta \phi$ which  becomes negligible over time. Since  thermal effects in Eq. (\ref{q1}) act in accordance with physical wave numbers,  we use the relation between the physical wave number and comoving wave number given by $k_{phy} = \frac{kc}{a} = k$. Therefore, the mode $\phi(k_{c})$ should be thermalized at the physical scale $k$ in the time interval $\sim \frac{1}{H}$ to satisfy the thermalization condition which means that the memory term should vanish within the Hubble time i.e. $\frac{1}{H\tau}>1$. The freeze-out wave number $k_{F}$ is at the point where this condition first holds, which for a negligible mass term  is
\begin{align}
k_{F} = H\sqrt{\frac{3\mathcal{C}}{\mathcal{D}}}.
\end{align}
The power spectrum for scalar fluctuations is calculated in the same manner as in cold inflation
\begin{align}\label{t1}
\mathcal{P}_{\mathcal{R}} = \left(\frac{H}{\dot\phi}\right)^{2}(\delta \phi)^{2},
\end{align}
where scalar perturbation of the inflaton field is obtained through
\begin{align}\label{t2}
(\delta{\phi})^{2}  =\frac{k_{F}T}{2\pi^{2}}.
\end{align}
Combining Eqs. (\ref{t1}, \ref{t2}), the power spectrum for warm G-inflation can be expressed as
\begin{align}\label{qa}
\mathcal{P}_{\mathcal{R}} = \frac{H^{3}T}{2\pi^{2}\dot\phi^{2}} \sqrt{\frac{3\mathcal{C}}{\mathcal{D}}}.
\end{align}
Based on calculations in \cite{Growing mode, Berera}, the form of the power spectrum in the high dissipation regime will be modified by a growing mode due to the coupling between radiation and inflaton fileds through the temperature dependence part of the dissipation coefficient in the high dissipation regime. Since the power spectrum (\ref{qa}) has been obtained by neglecting such a coupling, this will result in an accurate expression in the weak dissipation rather than strong dissipation regime for $c\neq 0$. In this sense, we will examine the model in section $V$ in the weak dissipation regime in order to obtain very reliable results. However, the power spectrum is still reliable even in the high dissipation regime for $c=0$. We therefore work with power spectrum Eq. (\ref{qa}) in general without considering the weak dissipation regime. Also, the spectral index is given by
\begin{align}\label{wq}
\nonumber n_{s}-1 &\equiv \frac{d \ln \mathcal{P}_{\mathcal{R}}}{d \ln k}= \frac{\mathcal{\dot P}_{\mathcal{R}}}{H\mathcal{P}_{\mathcal{R}}} = 3 \frac{\dot H}{H^{2}} + \frac{\dot T}{H T} -2 \frac{\ddot \phi}{H \dot \phi}  \\&+ \frac{1}{2}  \frac{\mathcal{\dot C}}{H \mathcal{C}} -  \frac{1}{2} \frac{\mathcal{\dot D}}{H \mathcal{D}} = -3 {\epsilon}_{H} + \delta_{T} - 2\delta_{\phi} + \frac{1}{2}  \delta_{\mathcal{C}} - \frac{1}{2}\delta_{\mathcal{D}},
\end{align}
with
\begin{align}
\nonumber  \delta_{\mathcal{C}} &\equiv \frac{d \ln \mathcal{C}}{d N} = \frac{\mathcal{\dot C}}{H \mathcal{C}} =\mathcal{C}^{-1} \left(\delta_{\Gamma} +  {\epsilon}_{H}+ \kappa_{X} \lambda_{X}\sqrt{\delta_{X}} \right.\\ & \left.\nonumber + 6 \frac{\delta_{GX}}{\delta_{X}} \left(2 \delta_{\phi} (\kappa_{X} + \lambda_{XX} - 2 \kappa_{X}^{2}) + 2 \lambda_{XX} \frac{\delta_{G\phi}}{\delta_{X}}\right)  \right.\\ & \left.+ 6 (1+ \kappa_{X}) (\eta_{GX}- \eta_{X}) \frac{\delta_{GX}}{\delta_{X}}\right),
\end{align}
and
\begin{align}
 \delta_{\mathcal{D}} \equiv \frac{d \ln \mathcal{D}}{d N} = \frac{\mathcal{\dot D}}{H\mathcal{D}}= 4 \mathcal{D}^{-1}(\eta_{GX} - \eta_{X}) \frac{\delta_{GX}}{\delta_{X}},
\end{align}
where we have used the following expression
\begin{align}
\nonumber\frac{\dot \kappa_{X}}{H}& = 2 \delta_{\phi} \left(\kappa_{X}- 2\kappa_{X}^{2} + \lambda_{XX}\right) + 2 \lambda_{XX} \frac{\delta_{G\phi}}{\delta_{X}}\\& - \kappa_{X}\lambda_{X}\sqrt{2\delta_{X}}.
\end{align}
Yet again, we note that Eq. (\ref{wq}) reduces to Eq. (36) in \cite{Moss:2008yb} for $G=0$. The corresponding running of the spectral index is indeed given by
\begin{align}\label{wq1}
\nonumber \alpha_{s} & \equiv \frac{d \ln n_{s}}{d \ln k} \\&= -3 \epsilon_{H} \eta_{H} + \delta_{T} \eta_{T} - \delta_{\phi} \eta_{\phi} + \frac{1}{2} \delta_{\mathcal{C}} \eta_{\mathcal{C}} - \frac{1}{2}\delta_{\mathcal{D}} \eta_{\mathcal{D}},
\end{align}
where $\eta_{T}, \eta_{\phi}, \eta_{\mathcal{C}}$,$\eta_{\mathcal{D}}$, $\eta_{X}$ and $\eta_{GX}$ are systemically defined as
\begin{align}
\nonumber \eta_{T} &= \frac{\dot \delta_{T}}{H \delta_{T}}, \ \ \eta_{\phi} = \frac{\dot \delta_{\phi}}{H \delta_{\phi}}, \ \  \eta_{\mathcal{C}} = \frac{\dot\delta_{\mathcal{C}}}{H \delta_{\mathcal{C}}}, \\ \eta_{\mathcal{D}}&=\frac{\dot\delta_{\mathcal{D}}}{H \delta_{\mathcal{D}}}, \ \ \eta_{X} = \frac{\dot \delta_{X}}{H \delta_{X}}, \ \ \eta_{GX} = \frac{\dot \delta_{GX}}{H \delta_{GX}}.
\end{align}
Therefore, Eqs.(\ref{wq}, \ref{wq1}) indicate that $n_{s}-1$ is of order $\epsilon$ and $\alpha_{s}$ is of order $\epsilon^{2}$. This means that the spectral index is scale-invariant and that the size of spectral index variations is very small which coincides with observation qualitatively. The tensor perturbations do not couple to thermal background and therefore gravitational waves are merely generated by the quantum fluctuations as in conventional inflation
\begin{align}\label{p1}
\mathcal{P}_{T} = 2 M^{-2}_{pl} \left(\frac{H}{2\pi}\right)^{2}.
\end{align}
The corresponding spectral index of gravitational waves is expressed as
\begin{align}
n_{T} = -2 \epsilon_{H} = -2 \frac{\epsilon}{\mathcal{A}}.
\end{align}
Dividing Eq. (\ref{t1}) by Eq. (\ref{p1}), the tensor-to-scalar ratio is given by
\begin{align}
r = \frac{\mathcal{P}_{T}}{\mathcal{P}_{\mathcal{R}}} = \frac{H}{T} \frac{2 \epsilon \sqrt{\mathcal{D}}}{\sqrt{3\mathcal{C}}\mathcal{A}^{2}}.
\end{align}
It is now observed that the tensor-to-scalar ratio can be much smaller, thanks to both thermal effects and Galileon field effects if both are strong, which is another synergy of both effects. Considering the slow-roll condition $\epsilon< \mathcal{A}$, we find that the energy scale of inflation merely bounds from above as
\begin{align}
r< \frac{H}{T} \frac{2 \sqrt{\mathcal{D}}}{\sqrt{3\mathcal{C}}\mathcal{A}}.
\end{align}
Also, the consistency relation becomes
\begin{align}
r= - \frac{H}{T} \frac{ \sqrt{\mathcal{D}}}{\sqrt{3\mathcal{C}}\mathcal{A}} n_{T},
\end{align}
which is not a fixed relation as in standard G-inflation ($r = -8.7 n_{T}$). The radiation energy density and universal temperature has the Stefan-Boltzmann relationship $\rho_{\gamma} = \frac{\pi^{2}g_{\star}}{30} T^{4}$ and therefore, utilizing slow-roll equations and power spectrum relation we find
\begin{align}
\frac{T}{H} =  \left(\frac{45}{4\pi^{2}}\right)^{\frac{1}{3}} \left(\frac{Q}{g_{\star}\mathcal{P}_{\mathcal{
R}}}\right)^{\frac{1}{3}}\left(\frac{3\mathcal{C}}{\mathcal{D}}\right)^{\frac{1}{6}}.
\end{align}
In fact, $\mathcal{C}> \mathcal{D}$  ameliorates the ratio of $\frac{T}{H}$, thus, the thermal effect is more obvious and the case is opposite when we have $\mathcal{C}< \mathcal{D}$. The condition for  warm inflation ($T>H$) to occur can  be obtained by $Q> g_{\star}\mathcal{P}_{R} $. Taking $g_{\star} $ of order $10^{2}$ and $\mathcal{P}_{R}$ of order $10^{-9}$, we  deduce that very small amount of dissipation results in warm inflation. One evaluates the variation of the inflaton field for observable scales with $\Delta N \simeq 4$ corresponding to $1<l<100$ as follows
\begin{align}
\frac{\Delta \phi}{M_{pl}} = \frac{\dot\phi \Delta N}{M_{pl} H} \simeq 5.2  \left(\frac{T}{H}\right)^{\frac{1}{2}}  \left(\frac{\mathcal{C}}{\mathcal{D}}\right)^{\frac{1}{4}} r^{\frac{1}{2}}.
\end{align}
In effect, it is possible to have large excursion of the inflaton field in a strong regime even if the tensor-to-scalar ratio is unobservable which can cure overlarge amplitudes of the inflaton field in conventional inflation. This is a striking characteristic of warm inflationary scenarios even in a non-G inflation limit.
\section{Warm Higgs G-inflation in weak dissipation and G-dominant regime ($Q \ll 1+ 3 \frac{\delta_{GX}}{\delta_{X}}$ \& $3 \frac{\delta_{GX}}{\delta_{X}}\gg 1$)}\label{5}
Now is the time to test the model at hand against observational data for particular forms of $V(\phi, T)$, $\Gamma(\phi, T)$ and $G(\phi, X)$. We take a general possible form of the generalized Galileon interaction term as \cite{Ohashi:2012wf}
\begin{align}\label{ff1}
G(\phi, X) = -\frac{\phi^{2p+1} X^{q}}{M^{4q+2p}},
\end{align}
where $M$ is a constant with dimension of mass. The aim of considering such general form is to show that for $(p, q) = (0,1)$, the simplest model of warm Higgs G-inflation is consistent with Planck results even for large self-coupling while the Higgs G-inflation which has been investigated in \cite{Ohashi:2012wf} could not exhibit such property since it may not be reheated for large self-couplings. Therefore, we compare our theoretical predictions against Planck likelihood including TT, TE and EE polarizations and BAO data \cite{Ade:2015xua} to confirm the consistency of the model with observations.

Upon considering (\ref{ff1}), the modified Klein-Gordon equation in weak dissipation $Q\ll1+3\frac{\delta_{GX}}{\delta_{X}} $ and G-dominant regimes $ 3 \frac{\delta_{GX}}{\delta_{X}}\gg 1$ reduces to
\begin{align}\label{as1}
-9qH^{2}\dot\phi^{2}\frac{\phi^{2p+1}X^{q-1}}{M^{4q+2p}} + V_{,\phi} \simeq 0,
\end{align}
where it deserves to be pointed out that the weak dissipation condition reduces to $Q\ll3\frac{\delta_{GX}}{\delta_{X}}$ using the G-dominant condition.

\begin{figure}[h]
\includegraphics[scale=0.65]{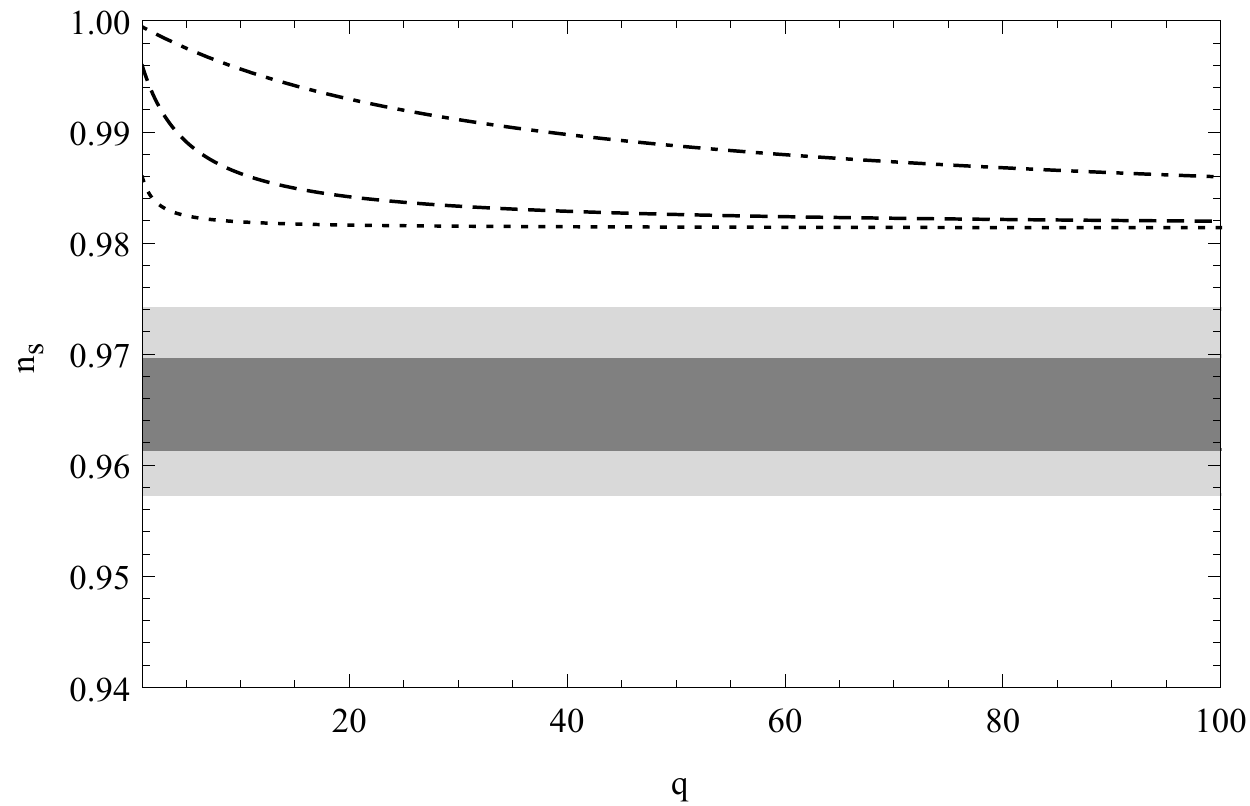}
\caption{The variation of spectral index $n_{s}$ versus $q$ for $N= 35$, $m=3$, $n=4$ and $\lambda = 0.13$ in which dotted, dashed and dot-dashed curves denote $p =0, 10$ and $100$, respectively. Also dark and light shades represent the range of tilt spectral index for $1 \sigma$ and $2 \sigma$ of Planck likelihood+ TTTEEE+ BAO.}
\label{fig1}
\end{figure}

Now, the chaotic inflation is characterized by the following power-law potential form
\begin{align}\label{akj2}
V(\phi) = \frac{\lambda}{n} M^{4}_{pl}\left(\frac{\phi}{M_{pl}}\right)^{n},
\end{align}
and from first principles in quantum field theory, dissipation coefficient has the following general form \cite{general dissipation}
\begin{align}\label{as3}
\Gamma(\phi, T) = \Gamma_{0} M_{pl} \left(\frac{\phi}{M_{pl}}\right)^{1-m} \left(\frac{T}{M_{pl}}\right)^{m} ,
\end{align}
where $\Gamma_{0}$ is connected to dissipative microscopic dynamics. The above expression for $m=-1$ becomes $\Gamma = \Gamma_{0}\frac{\phi^{2}}{T}$ which corresponds to a dissipative coefficient in non-SUSY case, $m=0$ gives $\Gamma = \Gamma_{0} \phi$ which corresponds to a SUSY case with an exponentially decaying propagator, $m=1$ gives $\Gamma = \Gamma_{0} T$, \cite{GammaT}, which corresponds to a high temperature SUSY case and $m=3$ gives $\Gamma = \Gamma_{0} \frac{T^{3}}{\phi^{2}}$, \cite{Bartrum}, which corresponds to a low temperature SUSY case.
Considering (\ref{as1},\ref{akj2}), one can obtain the inflaton velocity versus inflaton field as follows
\begin{align}\label{wa}
\dot\phi \simeq - \sqrt{2 } M_{pl}^{2} \zeta^{\frac{1}{2q}} \left(\frac{\phi}{M_{pl}}\right)^{-\frac{p+1}{q}},
\end{align}
with
\begin{align}
\zeta = \frac{n}{6q}\left(\frac{M}{M_{pl}}\right)^{4q+2p},
\end{align}
where we have considered a negative sign for the velocity field in order to have $\delta_{GX}>0$, therefore
\begin{align}\label{wb}
\mathcal{A} \simeq 3\frac{\delta_{GX}}{\delta_{X}} = \sqrt{\frac{n \lambda}{6}} \zeta^{-\frac{1}{2q}} \left(\frac{\phi}{M_{pl}}\right)^{\frac{nq+2p-2q+2}{2q}},
\end{align}
and using (\ref{e1},\ref{e2},\ref{wa},\ref{wb}), one can derive the following relation between temperature and inflaton field
\begin{align}\label{sz}
\frac{T}{M_{pl}} = \gamma_0~\zeta^{\frac{1}{q(4-m)}}\left(\frac{\phi}{M_{pl}}\right)^{\frac{2q-2mq-qn-4p-4}{2q(4-m)}},
\end{align}
where
\begin{align}
\gamma_0 = \left(\frac{60\Gamma_{0}\sqrt{n}}{\pi^{2} g_{*}\sqrt{3\lambda}}\right)^{\frac{1}{4-m}}.
\end{align}
Utilizing Eq. (\ref{sz}), the condition on $c$ can be written in terms of parameters of the model as follows
\begin{align}\label{dc}
|c| = \left|\frac{q (8-8m-nm)-4m(p+1)}{q(2-2m-n)-4(p+1)}\right| < 4.
\end{align}
Therefore, condition (\ref{dc}) should be checked for any special $p, q, m$ and $n$. Using the condition $\epsilon_{H} = 1$, one may find the value of the inflaton filed at the end of inflation and by inserting that into Eq.(\ref{vv}), one may obtain the following relation between the inflaton field at Hubble crossing time and the e-folding number
\begin{align}\label{dd}
\frac{\phi_{hc}}{M_{pl}} =\left(\frac{n}{6q}\right)^{\frac{1}{z_{0}}} \left(\frac{M}{M_{pl}}\right)^{\frac{4q+2p}{z_{0}}}\left(\sqrt{\frac{6n}{\lambda}}~\frac{z_{0}N+{nq}}{2q}\right)^{\frac{2q}{z_{0}}}.
\end{align}
Now, inserting (\ref{wa},\ref{sz}) into Eq. (\ref{qa}) by using $\kappa_{X} = q$ , Eq. (\ref{dd}) and definition for the spectral index, one may obtain the spectral index in terms of the e-folding number
\begin{align}\label{rt}
n_{s}-1 = - \frac{z_{1}}{z_{0}N+nq},
\end{align}
with
\begin{align*}
z_{0} &=nq+2q+2p+2,\\
z_{1} &=\frac{ nq(11-3m) + 2q(1-m) +(4p+4) (3-m)}{(4-m)}.
\end{align*}
Therefore, the model with $m=-1,0,1$ is just red tilted but for $m=3$ can be blue tilted for $n<2$,  red tilted for $n>2$ and gives $n_{s}=1$ for $n=2$. Furthermore, by inserting $m=3$ in Eq. (\ref{rt}) one can find that $n_{s}$ is very near unity, therefore, it is outside of the Planck data even for lowest possible e-folding number $N= 35$ \cite{Lowest1, Lowest2} as has been illustrated in figure (\ref{fig1}).
\begin{widetext}

\begin{table}[H]\begin{center}
\begin{tabular}{ |l|l|l|l|l|}
\hline
$ \ \ \ \ \ (p,q)   $ &$\ \ \ \ \ \ \ \ (0,1)$&$\ \ \ \ \ \ \ \ (0,2)$&$\ \ \ \ \ \ \ \ (1,1)$&$\ \ \ \ \ \ \ \ (3,2)$\\
 \hline
 $\ |{\delta_{G\phi}}/{3\delta_{GX}}|  $&$\ \ \ \ \ \ \ 0.0008$ &$\ \ \ \ \ \ \ 0.00042$& $\ \ \ \ \ \ \ 0.0018$ &$\ \ \ \ \ \ \ 0.0023$ \\
 \hline
 $\ \ \ \ \ \ x_{1}$& $\ \ -6.5<x_{1}<13.9$ &$\ \ -6.7<x_{1}<13.9$&$\ \ -6.5<x_{1}<14.2$&$\ \ -6.7<x_{1}<14.4$ \\
 \hline
$\ \ \ \ \ \ x_{2}$ &$\ -35.7<x_{2}<-15.2$  &$-35.9<x_{2}<-15.2$&$-36.4<x_{2}<-15.6$& $-37<x_{2}<-15.9$ \\
 \hline
$\ \ \ \ \ \ x_{3}$& $-12.2<x_{3}<-7.1$ &$-12.1<x_{3}<-6.9$&$-12.5<x_{3}<-7.3$&$-12.5<x_{3}<-7.2$   \\
 \hline
$\ \ \ \ \ \ x_{4}$ &$\ \ \ 0<x_{4}<10.2$  &$\ \ \ 0<x_{4}<10.3$&$ \ \ \ \ 0<x_{4}<10.4$&$\ \ \ \ 0<x_{4}<10.5$  \\
 \hline
$\ \ \ \ \ \ x_{5}$ &$\ \ 13.1<x_{5}<23.4$  &$\ \ 13.1<x_{5}<23.5$ &$\ \ \ \ 13.4<x_{5}<23.8$& $\ \ 13.6<x_{5}<24.1$\\
 \hline
$\ \ \  \ \ \ x_{6}$  & $\ \ -20.5<x_{6}<0$ &$\ \ -20.0<x_{6}<0$ &$\ \ \ -20.8<x_{6}<0$& $\ \ -21.1<x_{6}<0$\\
 \hline
$\ \ \ |c|<4$ & $\ \  \ \ \ \ \ \ m=0,1$ &$\ \ \ \ \ \ \ \ m=0,1$ &$\ \ \ \ m=-1,0,1$& $\ \ \ \ \ m=-1,0,1$ \\
 \hline
\end{tabular}
\caption{Constraints on the parameters of the model for different $ p , q$, $N=50$, $n=4$, $g_{*} = 100$ and $\lambda = 0.13$. Here we defined  $x_{1} = Log_{10} \left(\Gamma_{0}\right)$, $x_{2} = Log_{10}\left( r\right)$, $x_{3} = Log_{10}\left(\frac{\Delta \phi}{M_{pl}}\right)$, $x_{4} = Log_{10}\left(\frac{T}{H}\right)$, $x_{5} = Log_{10}\left(\mathcal{A}\right)$ and $x_{6} = Log_{10}\left(\frac{Q}{\mathcal{A}}\right)$ for convenience.}
\label{tab1}
\end{center}\end{table}
\end{widetext}

One may derive observable parameters of the model in terms of the spectral index as follows
\begin{align}\label{rr2}
&\mathcal{P}_R=\sqrt{\frac{q \lambda^3}{6 n^3}}\frac{\gamma_0}{2\pi^2}\zeta^{\frac{(5-m)(n-2)}{2z_0(4-m)}}\left(\sqrt{\frac{3n}{2\lambda}}\frac{z_1}{q(1-n_s)}\right)^{\frac{z_1}{z_0}},\\\notag
\\\label{tt1}
&r=\frac{\lambda}{6n\pi^2\mathcal{P}_R} \zeta^{\frac{n}{z_{0}}}\left(\sqrt{\frac{3n}{2\lambda}}\frac{z_1}{q(1-n_s)}\right)^{\frac{2nq}{z_0}},\\\notag
\\\label{tt2}
&\frac{T}{H} =\sqrt{\frac2q}\frac{12 n^2 \pi^2 \mathcal{P}_R}{\lambda^2} \zeta^{\frac{2-n}{z_{0}}}\left(\sqrt{\frac{2\lambda}{3n}}\frac{q (1-n_s)}{z_1}\right)^{\frac{4(nq+p+1)}{z_0}},\\\notag
\\\label{tt3}
&\left|\frac{\Delta \phi}{M_{pl}}\right| = \sqrt{\frac{96n}{\lambda}}\zeta^{\frac{1}{z_{0}}}\left(\sqrt{\frac{2\lambda}{3n}}\frac{q}{z_1}(1-n_s)\right)^{\frac{nq+2p+2}{z_0}},\\\notag
\\\label{tt4}
& Q =  \left(\sqrt{\frac{4n^{3}\pi^{5}\sqrt{{15g_{*}}} }{5q \lambda^{3}}}{ \mathcal{P}_{\mathcal{R}}}\right)^{4} \zeta^{\frac{3(2-n)}{z_{0}}}\nonumber
\\& ~~~~~~\times \left(\sqrt{\frac{3n}{2\lambda}}\frac{z_1}{q(1-n_s)}\right)^{-\frac{12(nq+p+1)}{z_{0}}},\\\notag
\\\label{tt5}
&3\frac{\delta_{GX}}{\delta_{X}} = \sqrt{\frac{n \lambda}{6}} \zeta^{-\frac{2}{z_{0}}}\left(\sqrt{\frac{3n}{2\lambda}}\frac{z_1}{q(1-n_s)}\right)^{\frac{nq+2p-2q+2}{z_{0}}},\\\notag
\\\label{tt6}
&\frac{\delta_{G\phi}}{3\delta_{GX}} = - \frac{2p+1}{3z_{1}}(1-n_{s}).\\\notag
\end{align}

Using the condition for  warm $T>H$,  weak dissipation  $Q<3\frac{\delta_{GX}}{\delta_{X}}$ and G-dominant $3\frac{\delta_{GX}}{\delta_{X}}>1$ regime, we can obtain a range for the value of $M$
\begin{align}
M_{min}<{M}< min\{M_{1}, M_{2}\},
\end{align}
where
\begin{align}\label{as}
\notag M_{min}& =M_{pl}\left(\frac{6q}{n}\right)^{\frac{1}{4q+2p}} \left(\sqrt{\frac{4n^{3}\pi^{5}\sqrt{15g_{*}}}{5q \lambda^{3}}}{ \mathcal{P}_{\mathcal{R}}}\right)^{\frac{4z_{0}}{(3n-8)(4q+2p)}}\\& \times \left(\frac{6}{n \lambda}\right)^{\frac{z_{0}}{6n-16}}\left(\sqrt{\frac{2\lambda}{3n}}\frac{2q (1-n_s)}{z_1}\right)^{\frac{13nq+14p-2q+14}{(3n-8)(4q+2p)}},\\\notag
\\
\notag M_{1} &= M_{pl}\left(\frac{6q}{n}\right)^{\frac{1}{4q+2p}}\left(\frac{6}{n \lambda}\right)^{\frac{z_{0}}{4(4q+2p)}} \\& \times \left(\sqrt{\frac{2\lambda}{3n}}\frac{2q (1-n_s)}{z_1}\right)^{-\frac{nq+2p-2q+2}{2(4q+2p)}},\\\notag
\\
M_{2} &=  M_{pl}\left(\frac{6q}{n}\right)^{\frac1{4q+2p}}\left(\sqrt{\frac2q}\frac{12 n^2 \pi^2 \mathcal{P}_R}{\lambda^2 }\right)^{\frac{z_0}{(n-2)(4q+2p)}} \notag\\
&\times\left(\sqrt{\frac{2\lambda}{3n}}\frac{2q (1-n_s)}{z_1}\right)^{\frac{4(nq+p+1)}{(n-2)(4q+2p)}}.\label{as2}
\end{align}
\begin{figure}[h]
\includegraphics[scale=0.65]{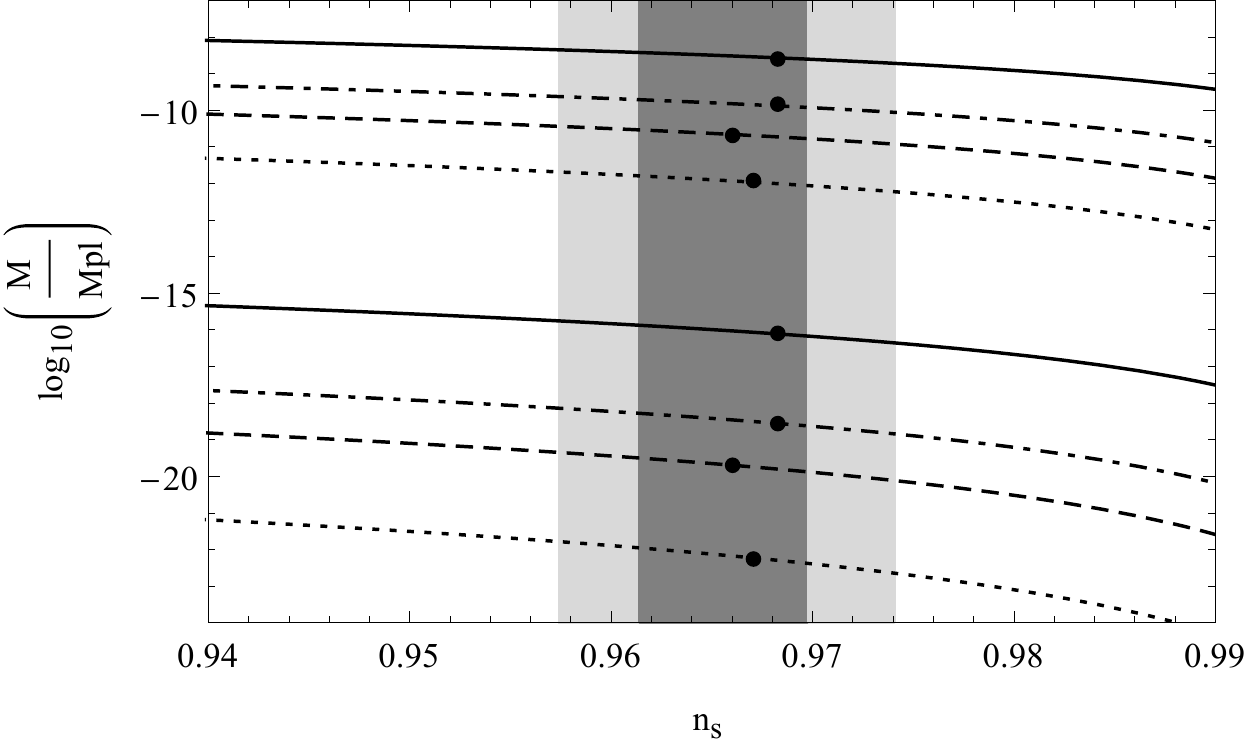}
\caption{The allowed region for $M$ and spectral index $n_{s}$ for $m=1$, $n=4$, $g_{*} = 100$ and $\lambda = 0.13$ in which dotted, dashed, dotdashed and solid curves denote lower and upper bounds on $M$ for $(p,q) = (0,1), (0,2), (1,1)$ and $(3,2)$, respectively, as well as dark and light shades represent the range of tilt spectral index for $1 \sigma$ and $2 \sigma$ of Planck likelihood+ TTTEEE+ BAO. Indeed, the points have been plotted for $N = 50$ and used Planck normalization $\mathcal{P}_{\mathcal{R}} = 2.44\times 10^{-9}$.}
\label{fig2}
\end{figure}

\begin{figure}[h]
\includegraphics[scale=0.65]{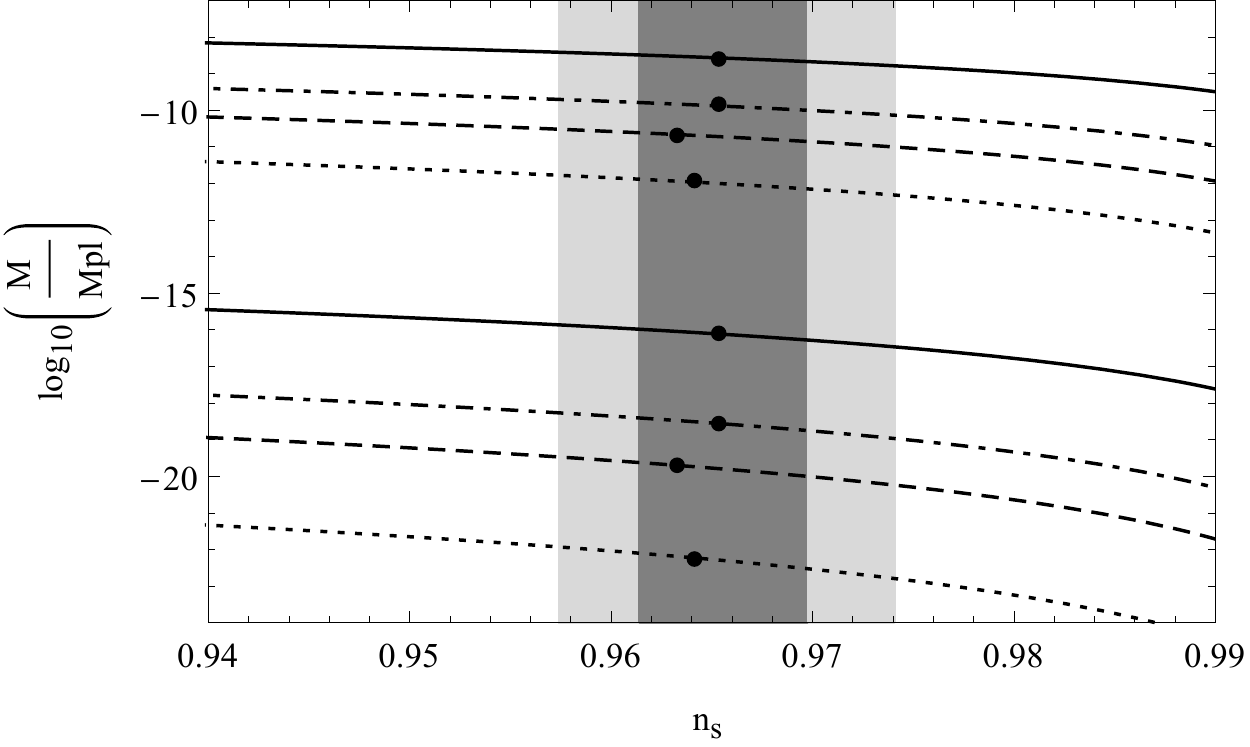}
\caption{The allowed region for $M$ and spectral index $n_{s}$ for $m=0$. Other information is the same with figure \ref{fig2}}
\label{fig3}
\end{figure}

\begin{figure}[h]
\includegraphics[scale=0.65]{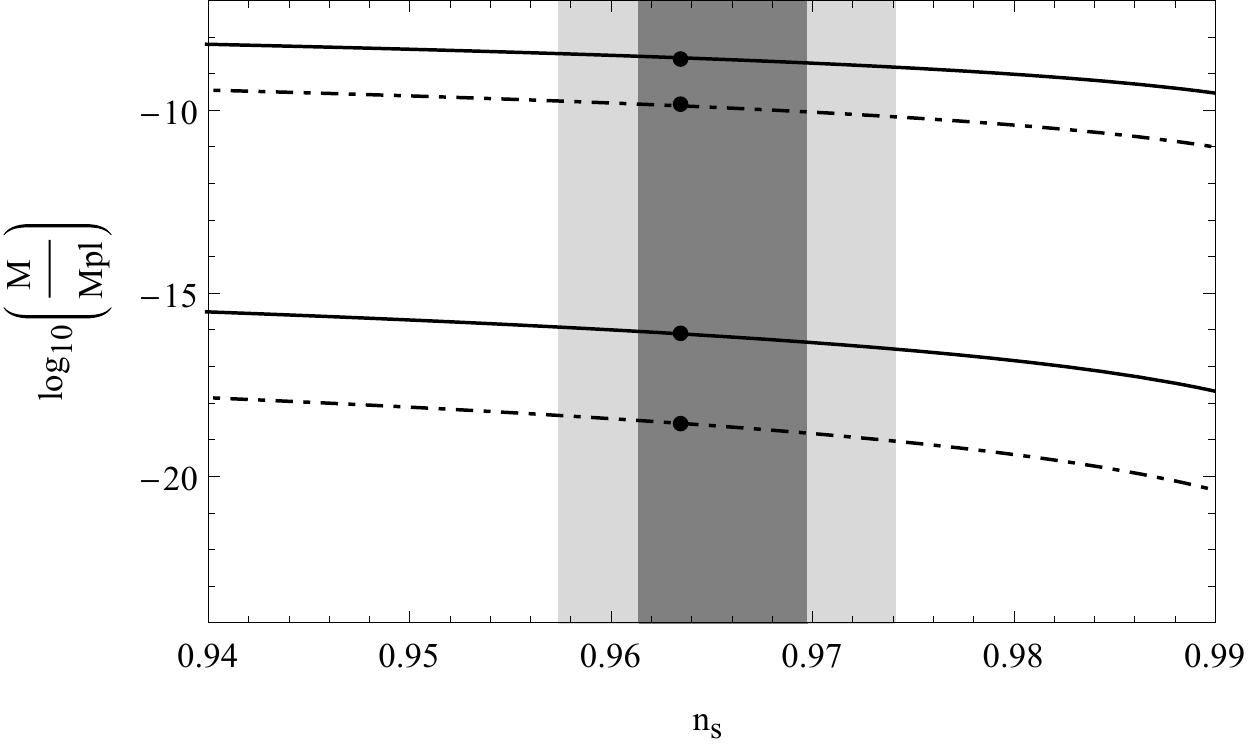}
\caption{The allowed region for $M$ and spectral index $n_{s}$ for $m=-1$ in which dotdashed and solid curves denote lower and upper bounds on $M$ for $(p,q) = (1,1)$ and $(3,2)$, respectively. Other information is the same with figure \ref{fig2}.}
\label{fig4}
\end{figure}
Now,  using Eq. (\ref{rr2}) and the bound on the value of $M$, we can also find the range of the parameter $\Gamma_{0}$. Having a look at Eqs. (\ref{tt1} -\ref{tt6}) and using Eq. (\ref{rt}) one may find that these expressions are independent of $m$ for particular values of the e-folding, and hence the range of these parameters for different $p$ and $q$ obtained in the table \ref{tab1}. We have also plotted the allowed region for $M$ versus spectral index in order to check if it is inside the Planck likelihood for $m=1, 0$ and $m=-1$ in figures \ref{fig2}, \ref{fig3} and \ref{fig4}. In fact, these models are well-consistent with observational date even for large self-coupling for very small values of $M$ as we expected from Eqs. (\ref{as}-\ref{as2}) since they are proportional to $\mathcal{P}_{\mathcal{R}}$ and $(1-n_{s})$ which have very small values. We also note that the value of $M$ will become larger by increasing the value of $p$ and $q$ as we can observe from figures 2, 3 and 4.

\section{Concluding remarks}\label{6}
In a warm inflationary scenario one avoids the reheating phase by introducing a dissipating exchange between  inflaton and radiation fields.  The disability of a Galileon-driven inflationary model, the G-inflation, to properly exhibit reheating motivated us to reconstruct G-inflation in the context of a warm scenario. In this sense, we modified the action by adding a generalized Galileon scalar field interaction where dynamical equation of the inflaton field is modified by the appearance of a viscose term. This resulted in  broader slow-roll conditions due to the synergy of thermal and Galileon effects and was achieved by applying a stability analysis to the resulting dynamical system. Indeed we obtained a novel but still scale-invariant form for power spectrum and found that the energy scale during horizon crossing is depressed by the synergy of the two effects.  Furthermore, the tensor-to-scalar ratio becomes substantially larger in weak warm G-inflation and insignificant in the opposite way.

Since we have not considered the coupling between the inflaton and radiation fields due to temperature dependent part of the dissipation coefficient ($c\neq 0$) in a high dissipation regime, the resulting power spectrum is more reliable in a weak dissipation regime. In this sense, we finally solved the model for the chaotic potential $\frac{\lambda}{n}\phi^{n}$  and $\Gamma(\phi, T) = \phi^{1-m}T^{m}$ with $G(\phi, X) = \phi^{2p+1}X^{q}$ and illustrated that Higgs G-inflation for a renormalizable potential ($n=4$), which may not otherwise be reheated for large self-coupling $\lambda \simeq 0.13$ in cold G-inflation, produces a scale-invariant power spectrum consistent with observations in the weak dissipation $Q\ll1+3 \frac{\delta_{GX}}{\delta_{X}}$ and G-dominant  $3 \frac{\delta_{GX}}{\delta_{X}}\gg1$ regime of the warm scenario for $m=-1, 0$ and $1$ and very small value of $M$. However, $m=3$ should be excluded since it cannot be inside the Planck likelihood for a sufficient e-folding number. Furthermore, we found that the value of $M$, for which the model is resistant with observation, will become larger by increasing the value of $p$ and $q$.

As the final remark, warm G-inflation with a low dissipative rate not only survives G-inflation, particularly Higgs inflation and its properties but also inherits the aforementioned properties coming from the warm scenario. We will focus attention on specific models and give the precise comparison with observations in a separate work.
As a matter of fact, natural inflation which suffers from the super-Planckian value of $f$ \cite{Ohashi:2012wf}, the parameter of the model, may also be cured in the warm scenario of G-inflation. Furthermore, we hope to be able to present an accurate analysis of the model in high dissipation for $c \neq 0$ in the near future. Also, the detailed issue of non-Gaussianity in the new scenario deserves more investigation.\vspace{2mm}\\

\noindent{\bf{Note added:}}
After completion of this work we became aware of  \cite{Herrera:2017qux} where the author has basically addressed the same issues. However, the present work goes much further by studying the consistency of the model and considerations of the general form of the  Galileon interaction term  $G(\phi, X)$, the general form of the dissipation coefficient and potential while taking into account temperature dependence, in contrast to what is done in \cite{Herrera:2017qux} where the author only considers the scalar field dependence. In addition, it should be mentioned that the freeze-out number which is derived in this work is different from that obtained in \cite{Herrera:2017qux} for $G(\phi, X) =  g(\phi) X$ due to the fact that to derive the freeze-out number, one should insert the full inflaton field which depends on the space and time, into covariant form of the modified Klein Gordon  equation (\ref{G1}) and, therefore, a Laplacian term will appear in the second order Langevin equation (\ref{G2}). As a result, the Laplacian term will be multiplied by a term `$\mathcal{D}$'  which changes the freeze-out number as compared to what has been obtained in \cite{Herrera:2017qux}.



\end{document}